\documentstyle[aps,prl,floats,epsf]{revtex}

\begin{document}

\draft

\wideabs{

%\preprint{ }

\title{A New Interpretation of Quantum Theory --- Time as Hidden Variable }
\author{{\normalsize Xiaodong Chen} \cite{chen} \\
{\normalsize Department of Physics, University of Utah, Salt Lake City,
UT 84112}}

\date{\today}
\maketitle

\begin{abstract} Using 2 more time variables as the quantum hidden
variables, we derive the equation of Dirac field under the principle
of classical physics, then we extend our method into the quantum
fields with arbitrary spin number. The spin of particle is shown
naturally as the topological property of 3-dimensional time +
3-dimensional space . One will find that the quantum physics of single
particle can be interpreted as the behavior of the single particle in
3+3 time-space .  
\end{abstract}

}

\narrowtext

Among all of the questions in quantum hidden variable theory, the
crucial one is : quantum physics is an elementary theory of physics, 
so the quantum hidden variables
should be very basic concepts of physics, then what are these variables ? 
The other important questions about quantum concepts are: i) Spin is
the basic property of particles and it is derived from quantum
physics, so is there a classical analogue of spin ? ii) The satisfactory
hidden variable theory should be a single particle theory, is that
possible for a single particle theory to recover the statistic
properties of quantum physics ? iii) Suppose we can get a single
particle theory under the principle of classical physics, how can we
interpret the non-local result of Bell's inequality ? iv) Why do we have
uncertainty relation? Finally, could the new interpretation give us more
knowledge than quantum physics does? There are many tries to answer
some of above questions as we can see in the work of Bohm \cite{bohm}
and Holland \cite{holland} . 
      
In this paper we will introduce a very interesting observation in
quantum hidden variable interpretation method: if we add two more time
variables $t_\theta$, $t_\phi$ as the quantum hidden variables,
i.e., 3-dimensional time 
(t, $t_{\theta}$, $t_{\phi}$) instead of 1 dimension, we'll 
find that the motion of single particle under 3+3 time-space posseses the same
qualitative behavior as the particle in quantum physics and the spin
of particle can be simply interpreted as the topological 
property  of 3-dimensional time. And then we will get the Dirac equation
and  Bargmann-Wigner equation.

Let us consider the two-slit interference experiment of electrons . We
know that in order to get the interference pattern, both slits should
have contributions to the result even if we control the electron to be
emitted one by one. That means if one tries to trace the
trajectory of a single electron, the only conclusion he can get is
that the
single electron  goes through
the two splits at the ``same time'', or in other word: ``a particle
can show at different places at the same time''. Let us express this word in
more detail : ``a particle can show at two different places at the same
time but this particle is still a single particle''. It sounds
weird, but there is one case that these things can really 
happen, that is: if and only if there are some hidden time variables
which we didn't find. That is to say, the electron passes through the
two splits at the same ``measure time'' but at different ``hidden time'' .       

Consider the case that the time in microcosm is more than 1
dimension ---
for  example 2-dimension, and we use clock to measure time. 
Since the clock can only measure the
1-dimensional time, i.e., the ``length'' of time, the
information about the ``direction'' of time (except the forward
and backward direction in the same line ) is unknown. Let us draw a circle
and build the polar
coordinate (t,$\theta$) on it, where t is modulus of time and $\theta$ is
angle . For the two points $t_1$(t,$\theta_1$),
$t_2$(t,$\theta_2$) on the circle with $r=t$, if we start our
measurement from the center $t_0$, because we can only
measure the ``length'' of time, we will treat $t_1$,$t_2$ as the same
point . Suppose there is a particle A. At time $t_1$, the
space coordinate of A is ($x_1$,0,0) and at time  $t_2$ 
the space coordinate of A  is ($x_2$,0,0), then using the knowledge of
1-dimensional time, we will conclude that a particle A can show at
different positions $x_1$, $x_2$ at the same time t. 

This situation is
similar to the situation of two-slit interference of electrons except that
the latter has the behavior of a plane wave. Now
let us make further explorations: what's the physical meanning of ``finding
A at position $x_0$'' in classic physics? That means at time t, 
our apparatus is at position $x_0$, and at the same time, A is 
also at position $x_0$, i.e., our apparatus meets the particle at position $x_0$
at time t. If 
the particle only shows at $x_0$ at time $t_1 \neq t$, and at the mean time,  our
apparatus only shows at $x_0$ at time t, we will miss that particle . 

Extend this logic to the case of 2-dimensional time. Then 
the meaning of finding a particle at position $x_0$ in 2-dimensional
time is : at time (t,$\theta_1$) our apparatus is at position
$x_0$, and at the same time (t,$\theta_1$) the particle also is at
position $x_0$, i.e., We meet the particle at $x_0$ at time
(t,$\theta_1$). If the 
particle only shows at $x_0$ at time (t,$\theta_2$) where $\theta_2
\neq \theta_1$, and our apparatus only shows at $x_0$ at time
(t,$\theta_1$), then we miss that particle . That is why even
though a particle can show at different places at the same time, we can
still only find one particle instead of 2 or more (Strictly speaking, 
here we need the Wave-Package Collapse postulate which will be
discussed later). Due to the lack of
information of the direction(the angle $\theta$) of time, we can only find the
particle by chance even though we have already known that the particle
will be at position $x_0$ at time t . The possibility of finding the
particle at position $x_0$ at time t depends on how many different
(t,$\theta)$ will show on position $x_0$ . If we 
let the total portion of that angle $\theta_i$( which passes through
$x_0$ at time t ) be divided by $2\pi$,  
and let the result correspond to the square of amplitude of the wave function of the
particle, then we will find that the measurement in the multi-dimensional
time of a single particle has the same statistic property as the
measurement in quantum physics. Furthermore, if a particle at different
$\theta_i$ with the same radius t has different energies, then at time t, if
we measure energy of the particle, we cannot be sure about which result we will
get. The possibility to find the energy which equals a particular E will
depend on the portion of total  $\theta_i$ which corresponds to the
same energy E divided by $2\pi$. 

We can go one step further. Obviously the above picture of
2-dimensional time has the non-local property: it does not satisfy the
causality of 1-dimensional time --- we can not link $t_1$ and $t_2$
which are on the same circle by the causal relation. The only thing
we can expect about causality is the distribution of total time vice
space, that is similar to the case in quantum physics, where we can only have the
causality for whole wave function. All these arguments can be easily
extended to 3- or more- dimension case. Moreover, since we have more
degrees of freedom
in time, then is that possible to have a rotation in time coordinate
which is just the same as spin ?  
              
Although the above picture is rather rough, it 
still shows the potential what the 2 or 3 dimensional time can
realize. First let us guess the dimension of time is 3, just for
symmetry reason, later we will find the strong support for this
guess. The rest questions are what is the behavior for a free
particle which moves in the 3+3 space-time and what is the mechanism in
that case?  Our job is to find how does the 3-dimensional time maps into 3
dimensional space for a free particle. Is that
just the plane wave function? Is it possible that the relations between
time and energy and between the position and momentum satisfy the
uncertainty relation? Is that possible that the mechanism of the motion of
the free electron in 3+3 space-time satisfies the Dirac equation?      

Imagine a ghost in microcosm who can watch the single particle
without interrupting it and suppose he uses the clock to measure time. He will find 
when the clock
points to time t, this particle shows at lots of different places with
different ``time angles'' $t_\theta$, $t_\phi$. Furthermore, he can
watch the evolution of the particle in each position when t changes,
i.e., he traces the changes of each position(at different $t_\theta$,
$t_\phi$) when t changes, then there will be lots of paths distributed
in space, each path may have different weights, and on each path,the
causality is satisfied. This picture is the 
same as the idea of Feynman path integral. That is, we can let each time path
from the center of time sphere to the surface correspond to each
Feynman path, and let the surface of time sphere correspond to the
surface of wave function.  In other aspect, in the whole
space the positions of the single particle form a field in which each
``force line'' passes through each different $t_\theta$, $t_\phi$ at
the specific time, so our question is: how to build a field on a spherical
surface $S^2$ in 3-dimensional time coordinate? 

This situation
is the same as the situation of Dirac monopole: the ``force line'' of
our field is just the Hopf bundle on $S^2$. It is well known in
monoploe theory \cite{Aitchison} that, according to the nontrivial global topological
property of $S^2$, the transition function for monopole bundle on spherical surface
is $exp(-2ig\phi)$ (here $\hbar = 1$), and it is not single-valued unless it satisfies 
the quantization condition:
\begin{equation}
e^{-4\pi ig} = 1 .
\end{equation}
Hence winding number g = 0, $\pm$1/2, $\pm$1,$\cdots$. We will find
that, in our case, g is just the spin of the single particle.
   
Consider a ``time'' sphere whose radius is 1/2 and  sperical coordinate on the
surface is (1/2, $t_\theta$, $t_\phi$), resting on a $u=x_3+is$ plane at
the south pole. Here $x_3$ is the 3rd coordinate of space, and s is the
combination of space coordinate x and y(each point on s axis
corresponds to a circle in $x_1$-$x_2$ plane). Let the axis $x_0$ which corresponds
to one specific time direction pass through
the south pole o and be perpendicular to $x_3$-s plane, then the whole
coordinate system is shown in Fig1. Let Z=
$\left(\begin{array}{c} z_1 \\ z_2
\end{array}\right)$ be the stereographic projection 
coordinates of a point in the northern hemisphere $U_n$ of $S^2$,
where $z_1$ =$x_0$, $z_2$ = $z+is$.  
on the unit sphere, then we have: 
\begin{equation}
x_0^2-x_1^2-x_2^2-x_3^2 = x_0^2-s^2-z^2 
             =\left | z_1 \right | ^2 - \left | z_2 \right | ^2 = 1 .
\end{equation}
Compare it with the condition in Hopf bundle \cite{Aitchison} $x_1^2+x_2^2+x_3^2+x_4^2
= 1$, the only difference is that the sign of $x_i$(i=1,2,3)  
becomes negative. This negative sign comes from the time metric just the same 
as the time in relativity. Then from (2) and Fig1., we find  $x_0 =
1/2(1+\cosh {t_\theta})$, $x_3 = 1/2 \sinh {t_\theta} \cos {t_\phi}$,
$s=1/2 \sinh {t_\theta} \sin {t_\phi}$, Z can also be written as:
\begin{equation}
Z= \cosh {t_\theta \over 2} \left(\begin{array}{c} \cosh {t_\theta
\over 2} \\ \sinh {t_\theta \over 2}e^{i t_\phi } \end{array}\right) .
\label{eq:2-row}
\end{equation}
Since Z is the stereographic projection coordinate, we can omit the
common coefficient $\cosh {t_\theta \over 2}$, then:
\begin{equation}
Z=\left(\begin{array}{c} \cosh {t_\theta \over 2} \\ \sinh {t_\theta
\over 2}e^{i t_\phi } \end{array}\right) .
\end{equation}
Rewrite our representation, let $z_1$=$\left(\begin{array}{c} x_0 \\ 0
\end{array}\right)$, 
$z_2$=$\left(\begin{array}{c} z \\ s \end{array}\right)$, then the
expression of Z can be written 
as 4-component form:
\begin{equation}
Z=\left(\begin{array}{c} \cosh {t_\theta \over 2} \\ 0 \\ \sinh
{t_\theta \over 2}cos{t_\phi } \\ 
\sinh {t_\theta \over 2}sin{t_\phi} \end{array}\right) .
\end{equation}
If we add the U(1) bundle to the surface, then Z becomes:
\begin{equation}
Z \rightarrow Z=\left(\begin{array}{c} \cosh {t_\theta \over 2}
\\ 0 \\ \sinh {t_\theta \over 2}cos{t_\phi } \\ 
\sinh {t_\theta \over 2}sin{t_\phi} \end{array}\right)e^{i\chi} .
\label{eq:4-row}
\end{equation} 
On the north hemisphere, we can let 
\begin{eqnarray}
\cosh {t_\theta} = \frac{1}{\sqrt{1-\frac{v^2}{c^2}}}  \\
\sinh{t_\theta} = \frac{\frac{v}{c}}{\sqrt{1-\frac{v^2}{c^2}}}  \\
cos{t_\phi} = v_3       \\
sin{t_\phi} = v_s 
\end{eqnarray}
where $v = \sqrt{v_{3}^{2}+v_{s}^{2}}$.
Then, we make a substitution:
\begin{equation}
v_s \rightarrow v_1 + iv_2
\label{eq:sub1}
\end{equation}
where $v_1$,$v_2$,$v_3$ are components of velocity of free particle in space
coordinate, then the Eq.~\ref{eq:4-row} becomes
\begin{equation}
 Z = \cosh {t_\theta \over 2} \left(\begin{array}{c} 1 \\ 0 \\
 \frac{\sinh {t_\theta \over 2}cos{t_\phi }}{
 \cosh{t_\theta \over 2}} \\ \frac{\sinh {t_\theta
 \over 2}sin{t_\phi}}{\cosh{t_\theta \over 2}}\end{array}\right)e^{i\chi} 
\end{equation}
\begin{equation}
=  \sqrt{\frac{m+\frac{m}{\sqrt{1-\frac{v^2}{c^2}}}}{2m}}
\left(\begin{array}{c} 1 \\ 0 \\
\frac{mv_3}{m+\frac{m}{\sqrt{1-\frac{v^2}{c^2}}}} \\ \frac{mv_1+imv_2} 
{m+\frac{m}{\sqrt{1-\frac{v^2}{c^2}}}}\end{array}\right)e^{i\chi} .
\label{eq:Dirac}  
\end{equation}
If we let m be the static mass of particle, then $mv_i = p_i$(i=1,2,3) and $ 
\frac{m}{\sqrt{1-\frac{v^2}{c^2}}} = E$. Let $\chi = p^{\mu}x_{\mu}$ ($\mu$
= 0,1,2,3), we can find that Z
is just one of the solution to ``positive-energy'' of Dirac field in 
$\sigma_z$ representation(Here $\sigma$ is Pauli matrix.)
It is easy to see that, let $v_s = v_1 - iv_2$ and
interchange the 2nd row and 1st row, and then 4th row and 3rd row, we'll
 obtain the other solution with ``positive-energy''.  
On the south hemisphere, $x_0
=\frac{1-\cosh{t_\theta}}{2}=\sinh^2 {\frac{t_\theta}{2}}$, and let
$\chi \rightarrow \chi - t_\phi$, then Eq.~\ref{eq:2-row} becomes:  
\begin{equation}
Z=\sinh{t_\theta \over 2}\left(\begin{array}{c} \sinh {t_\theta \over
2}e^{-it_\phi } \\ \cosh{t_\theta \over 2} \end{array}\right) .
\end{equation}
Written as 4-components representation and let $v_s = -v_1-iv_2$, Z
will become the solution of 
``negative-energy'' of Dirac field in $\sigma_z$
representation. Similarly, if we let $v_s = -v_1 + iv_2$ and interchange
the 2nd row and 1st row, 4th row and 3rd row, then we will get the
other solution of ``negative-energy''. So we find that Z in the whole
sphere corresponds to 4 solutions of free Dirac field in $\sigma_z$
representation . Now we put Eq.~\ref{eq:4-row} into Dirac equation
\begin{equation}
(i\gamma ^\nu \partial_{\nu} - m)\psi = 0
\end{equation}
and multiply the both sides of the equation by
$Z^+\gamma^0$, then we get 
\begin{equation}
iZ^+\gamma^0 \gamma ^\nu \partial_{\nu}Z = m.
\end{equation}
Let us make a substitution 
\begin{equation}
\partial_{x_1} + \partial_{x_2} \rightarrow \partial_s
\label{eq:sub2}
\end{equation}
and suppose that in free particle case, $t_\theta, t_\phi$ don't depend on
$x_i$(i=0,1,2,3), then after simple calculation, we have 
\begin{equation}
\cosh{t_\theta}\partial_{x_0}\chi +\sinh{t_\theta} \cos{\phi} \partial_{x_3}\chi 
 + \sinh{t_\theta} \sin{\phi} \partial_{s}\chi = -im.
\label{eq:free}
\end{equation}
However, as we can see from Fig1., the coordinates of unite vector $\hat{n}(n_1,n_2,n_3)$ 
have the relation that: 
\begin{eqnarray} 
n_1 = \hat{n} \cdot \hat{x_3} = \sinh{t_\theta} \cos{\phi} \\
n_2 = \hat{n} \cdot \hat{s} = \sinh{t_\theta} \sin{\phi}  \\
n_3 = \hat{n} \cdot \hat{x_3} = \cosh{t_\theta}
\end{eqnarray}
where ``$\cdot$'' means the scalar product of two vector.
Then the Eq.~\ref{eq:free} can be written as :
\begin{equation}
- \hat{n} \cdot \nabla \chi = im
\end{equation}  
or 
\begin{equation}
\vec{Q} = - \nabla \chi = im\hat{n}
\end{equation}  
where $\nabla = \hat{x_0}\partial_{x_0} + \hat{x_3}\partial_{x_3} +
\hat{s}\partial_{s}$.   
Compare the above two equations with the case in static electric field
$\vec{E} = - \nabla \phi = \frac{q}{r^2} \hat{r}$, 
we find that $\vec{Q}$ has the form of classical field and the free Dirac field
corresponds to  
classical static field in 3-dimensional time-space, so we call $\vec{Q}$ 
``time field''.  

Here, we use the Dirac field in $\sigma_z$ representation which corresponds
to a particular coordinate system $x_3-s$ in Fig1. The s axis can be
treated as the 
combination of $x_1,x_2$, and we make such substitution as
Eq.~\ref{eq:sub1} and Eq.~\ref{eq:sub2} since each point on s axis
corresponds to a circle in 
$x_1$-$x_2$ plane. In general, the mapping from 3-dimensional space to
time spherical surface  is completed by 2 steps. First, from the spacial $S^3 \cong SU(2)
\rightarrow $ spacial $S^2$, and 
then from the spacial $S^2$ to time spherical surface $S^2$. Instead
of Eq.~\ref{eq:Dirac},  in 
north hemisphere, Z can be written as:
\begin{equation} 
 Z = \sqrt{\frac{m+\frac{m}{\sqrt{1-\frac{v^2}{c^2}}}}{2m}}
     \left(\begin{array}{c} 1 \\ 0 \\   
     \frac{m\vec{v} \cdot \vec{\sigma}}{m+\frac{m}{\sqrt{1-\frac{v^2}{c^2}}}}
       \left( \begin{array}{c} 1 \\ 0 \end{array} \right)
\end{array}\right)e^{i\chi}.
\end{equation}
This is the ``positive energy'' solution of Dirac equation of free
electron in any representation. The same method can be used in the
other 3 cases. 

Refer to monopole theory. Here Z corresponds to Hopf bundle with
winding number g = 1/2 in monople theory.  It's easy to see that if we choose the
representation Z as the direct product representation of $Z_1,Z_2$,i.e.,
$Z= Z_1Z_2$, where $Z_i$(i=1,2) are 4-components vectors as eq.(5),
then the new Z corresponds to Hopf bundle with winding number
g=1. Generally, $Z=\prod\limits_{i}^{n} Z_i$ (i=1..n) corresponds to
Hopf bundle with 
g=n/2 . Since each $Z_i$ satisfies eq.(16), the whole Z is just the
spinor which satisfies Bargmann-Wigner equation, and the winding
number g just correspond to spin in quantum field theory!

From all the above, we can see that spin can be naturally obtained from the
topological property of the particle in 3+3 dimensional time-space, 
but if time dimension is less than or more than 3, we cannot get
the same topological result. That is why we think time is
3-dimensional. Also, we obtain 3 interesting observations: 

i) The $E > 0$ solution corresponds to the Hopf bundle in north
hemisphere, and $E < 0$ solution corresponds to the Hopf bundle in south
hemisphere. That means the time direction of electron with negative
energy is opposite to the time direction of electron with positive
energy, so one can guess that the negative energy corresponds to the
electron's past. The reason that the ``negative sea'' is fully occupied
is because the history of electron is fully occupied. So in our case, the
Dirac equation is a single particle equation.

ii) If m = 0, we should modify our method to get the field
equation. We need move our original point o and the whole  
$x_3-s$ plane of Fig1. along $x_0$ axis 1/2 upward, i.e., move the
original point of our coordinate to the center of time sphere, then
the whole system has spherical symmetry. This symmetry corresponds to
gauge invariant. When $m \not= 0 $, we lose spherical symmetry,
i.e., we should choose one particular direction as our $x_0$ axis, and
one particular direction as our $\hat{n}$ direction. This case
corresponds to non-gauge invariant case. So this mechanism is just
the same as the symmetry spontaneous symmetry violation in Higgs
mechanism. 

iii) In 3+3 dimensional time-space, the most surprised thing is 
that, even if two particles stay at the same position (x,y,z) and at
the same time t, it is possible that they cannot ``see'' each other, in
another word, they have no interaction with each other --- if the
coordinates of particle 
1 is $(t,t_{\theta_1}, t_{\phi_1},x,y,z)$, but of particle 2 is
$(t,t_{\theta_2}, t_{\phi_2},x,y,z)$. This can interpret why we have
the Bose-Einstein Condensate and Superconductivity: when the
distributions of time angles of all the particles are fit very well
such that, within each small space area, and at time t, no two
particles have same time angle $t_{\theta}, t_{\phi}$, then no two
particles can ``see'' each other. These kinds of system  will not
contain any interaction.       

Furthermore, we have found that the solutions of
quantum field equation of free particle correspond to Hopf bundles in
monopole theory, and each  Hopf fiber corresponds to each plane wave with
different momentum $\vec{p}$. Consider two extreme cases. When the space
position of particle is confined at x at time t 
--- the particle stays at x in all the time angles
$(t,t_{\theta},t_{\phi})$, and each angle in time surface corresponds
to one Hope fiber and each Hope fiber corresponds to each different
$\vec{p} $, so the different fiber corresponds to each path of particle with 
different speed. Then after time t, The particle at same t but with different 
time angles reaches the different positions. This picture conforms to
wave-package diffusion in quantum theory; when the
particle is in  fixed momentum $\vec{p}$, each space point x can
only contain one Hopf bundle(i.e., one time angle), and all different time
angles will be distributed to the whole space, but with the same Hopf
bundle, we can find that particle everywhere. This picture describes the 
uncertainty relation in 3+3 dimensional time-space. 

In addition, we should also add
``Wave-Package-Collapse'' conjecture to our theory. In our case, that
is: if we try to measure some physical attributes of the particle, but by
chance, the apparatus only meets a portion of time angle of the
particle(suppose the interaction between the apparatus and the
particle  will change the distribution of 
time angles of that particle in the whole space.), then after the
measurement, the distribution of time angles of the particle will
depend on the portion of time angles which we 
measured. The case is the same as a needle sticks into an inflated balloon,
the whole surface of balloon will be destroyed. 

The detail of the dynamical properties and the metric of 3+3 time-space will be
discussed in future. 

\clearpage
\begin{figure}[ht!]
\begin{center}
\epsfysize=5.0in
\epsfxsize=5.0in
\epsfbox{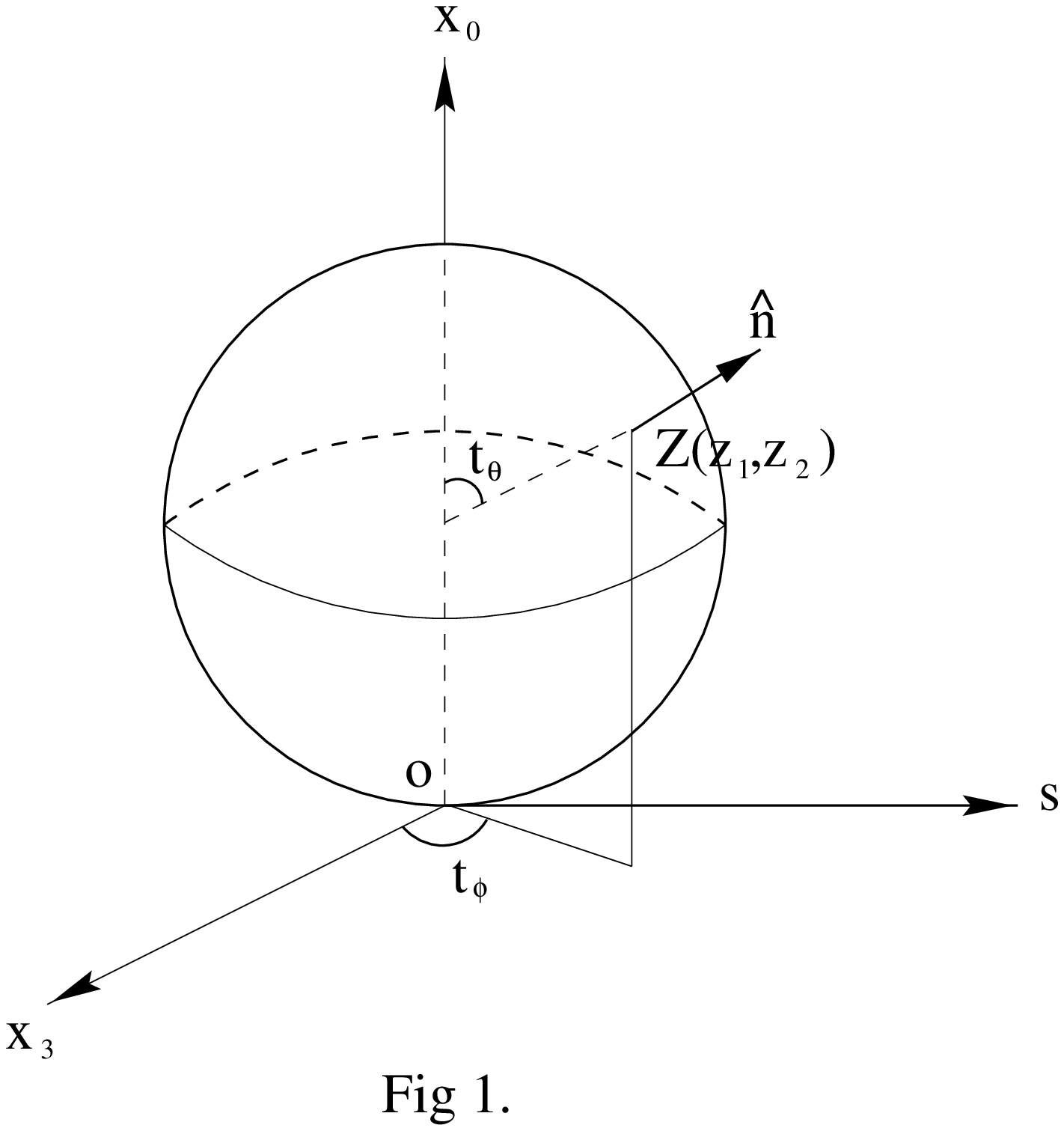}
\label{grid1}
\end{center}
\end{figure}

\end{document}